\documentclass[journal=nalefd,manuscript=letter, layout=twocolumn]{achemso}

\setkeys{acs}{articletitle = true}
\def\acs@type@default{letter}
\def\acs@type@list{letter}

\let\oldmaketitle\maketitle
\let\maketitle\relax

\usepackage{color}
\usepackage[version=4]{mhchem}
\usepackage{siunitx}

\DeclareSIUnit{\wtpercent}{wt.\%}
\DeclareSIUnit{\inhg}{inHg}
\newcommand{\ldr}{$\mathrm{LD}^{r}_{\mathrm{2D}}$}
\newcommand{\stwod}{$S_{\rm 2D}$}

\newcommand{\soneone}{$S_{11}$}

\newcommand{\ivv}{I_{\mathrm{VV}}}
\newcommand{\ivh}{I_{\mathrm{VH}}}
\newcommand{\ihh}{I_{\mathrm{HH}}}

\usepackage{hyperref}
\usepackage{graphicx}
\usepackage{dcolumn}
\usepackage{bm}

\author{Joshua S.~Walker}
\affiliation{Department of Physics, University of Wyoming, Laramie, WY, 82071}

\author{Jeffrey A.~Fagan}
\affiliation{National Institute of Standards and Technology, Materials Science and Engineering Division, Gaithersburg, MD, 20899}

\author{Adam J.~Biacchi}
\affiliation{National Institute of Standards and Technology, Nanoscale Device and Characterization Division, Gaithersburg, MD, 20899}

\author{Valerie A.~Kuehl}
\affiliation{Department of Chemistry, University of Wyoming, Laramie, WY, 82071}

\author{Thomas A.~Searles}
\affiliation{Department of Physics \& Astronomy, Howard University, Washington, DC, 20059}

\author{Angela~R.~Hight Walker}
\affiliation{National Institute of Standards and Technology, Nanoscale Device and Characterization Division, Gaithersburg, MD, 20899}

\author{William D.~Rice}%
\affiliation{Department of Physics, University of Wyoming, Laramie, WY, 82071}

\email{wrice2@uwyo.edu}

\title{Global alignment of solution-based single-wall carbon nanotube films via machine-vision controlled filtration}

\begin{document}

\twocolumn[
\begin{@twocolumnfalse}
\oldmaketitle
\begin{abstract}
Over the past decade, substantial progress has been made in the chemical processing (chiral enrichment, length sorting, handedness selectivity, and filling substance) of single-wall carbon nanotubes (SWCNTs).  Recently, it was shown that large, horizontally-aligned films can be created out of post-processed SWCNT solutions.  Here, we use machine-vision automation and parallelization to simultaneously produce globally-aligned SWCNT films using pressure-driven filtration.  Feedback control enables filtration to occur with a constant flow rate that not only improves the nematic ordering of the SWCNT films, but also provides the ability to align a wide range of SWCNT types and on a variety of nanoporous membranes using the same filtration parameters.
Using polarized optical spectroscopic techniques, we show that meniscus combing produces a two-dimensional radial SWCNT alignment on one side of the film.  After we flatten the meniscus through silanation, spatially-resolved nematicity maps on both sides of the SWCNT film reveal global alignment across the entire structure. From experiments changing ionic strength and membrane tribocharging, we provide evidence that the SWCNT alignment mechanism stems from an interplay of intertube interactions and ordered membrane charging.  This work opens up the possibility of creating globally-aligned SWCNT film structures for a new-generation of nanotube electronics and optical control elements.  
\end{abstract}

\end{@twocolumnfalse}
]

Significant interest in one-dimensional (1D) nanocrystals (NCs) follows from their highly anisotropic properties of electrical and thermal transport, optical absorption, radiative emission, and conduction. 
Typically, physical attributes observed in these NCs are significantly enhanced along the extended 1D crystal axis relative to the short axes, the latter often serving to impose strict quantum mechanical boundary conditions on the band structure.
Researchers have utilized the anisotropic nature of 1D NCs in polymer chains, liquid crystals, inorganic crystals, and carbonaceous ribbons (graphene, e.g.), and nanotubes to explore physically-interesting 1D behaviors like Luttinger liquids~\cite{VoitRepProgPhys1995, BockrathNature1999, ShiNaturePhotonics2015}, time-reversal-invariant Majorana chains~\cite{FidkowskiPRB2010, FidkowskiPRB2011}, Wigner crystals~\cite{DeshpandeNaturePhys2008}, ultrastrong light-matter coupling~\cite{GaoNaturePhotonics2018, HoPNAS2018}, high-harmonic generation~\cite{TorresPRB2004}, Aharonov-Bohm physics~\cite{AjikiPhysicaB1994, ZaricScience2004}, intersubband plasmons~\cite{YanagiNatureComm2018}, and topological insulators~\cite{AutesNatureMater2016}.
Additionally, 1D NCs have been envisioned in a wide range of technologically-important applications, such as high current-carrying capacity conductors~\cite{YaoPRL2000, HuangNanoLett2015}, rectifers~\cite{HarnackNanoLett2003}, far-infrared polarizers~\cite{RenNanoLett2009} and detectors~\cite{HeNanoLett2014}, gas and molecular sensing~\cite{CubukcuPNAS2009, WeiNanoLett2018}, flexible electronics~\cite{ChenNanoLett2011},  photoelectron emission~\cite{GreenNanoLett2019}, and directional heat transport~\cite{ZhangACSNano2018}.  
Consequently, enhancing these anisotropic effects by aligning 1D NCs along a common axis via easy-to-control mechanical, electrical, or magnetic external forces is highly desirable.
However, the high degree of van der Waals interactions per unit mass in these nanosystems promotes particle aggregation, which contributes to the difficulty in creating globally-aligned macroscopic films of 1D NCs.

Among the major 1D NC groupings, single-wall carbon nanotubes (SWCNTs) are particularly difficult to reproducibly align, especially after they have gone through solution-based processing.  
Despite the well-known challenges involved in nanotube preparation, strong interest remains in working with SWCNTs due to their unique band structures and exemplary physical properties~\cite{NanotSpringer2013}. 
Substantial research into chemically processing SWCNTs to achieve and enhance these superlative behaviors has produced significant breakthroughs in chiral and type enrichment~\cite{ZhengTopCurrentChem2017, BatiNanoscale2018, JanasMaterChemFront2018, FaganNanoscaleAdv2019}, length sorting~\cite{KhripinAnalChem2013}, tube filling with atoms and molecules~\cite{TakenobuNatureMater2003, CambreAngewChem2011, FaganACSNano2011, CambreNatureNano2015, CampoNanoHoriz2016}, and handedness selectivity~\cite{AoJACS2016, WeiNatureComm2016}.
Alignment of nanotubes along a preferred direction has also been achieved, but often with significant caveats or over a limited scope.  
Researchers have used a variety of techniques to align nanotubes including non-chiral-enriched, vertically-oriented SWCNT forests~\cite{HataScience2004, MurakamiChemPhysLett2004}, mechanical pulling of polymers~\cite{JinAPL1998}, electrostatic-enhanced dropcast films~\cite{LeMieuxScience2008}, magnetic alignment~\cite{WaltersChemPhysLett2001}, nanowire self assembly~\cite{HobbieACSNano2009}, and feedstock-driven growth~\cite{CubukcuPNAS2009, CheACSNano2012}.
Recently, a significant step forward in SWCNT alignment was taken when He et al.~\cite{HeNatureNano2016, GaoRoyalSociety2019} demonstrated that SWCNTs formed along a particular axis when a nanotube solution was slowly filtered through a hydrophilic, polyvinylpyrrolidone (PVP)-coated nanoporous membrane.
This observation allows researchers to produce well-aligned polarized SWCNT films \textit{after} solution-based chemical processing (e.g., chiral enrichment or length sorting).
Unfortunately, this technique is challenging to reproduce and even more difficult to scale up, which has hindered its widespread adoption.
Furthermore, previous research with and on aligned SWCNT films claim global alignment, a statement challenged by the absence of a true macroscopic characterization technique, as well as a complete reliance on single-side film measurements.

In this Letter, we use an automated and parallelized filtration system to reproducibly and simultaneously create multiple highly-aligned, solution-based SWCNT films, thus allowing us to explore a large set of chemical and physical parameters under well-controlled conditions.
Using machine vision, we both measure and control the filtration flow rate for different filter membrane pore sizes by monitoring the solution meniscus and regulating the transmembrane pressure.  
This automated feedback loop produces a constant filtration flow rate, which not only improves SWCNT alignment, but also enables this technique to be easily applied to different varieties of synthesized SWCNTs.
Additionally, we use a combination of polarized optical techniques and glass silanation to discover and remove the formation of a meniscus-created radial SWCNT alignment.   
Spatial mapping of both sides of the SWCNT film using polarized Raman scattering shows a two-dimensional nematic ordering parameter, \stwod, of $\approx$0.9 throughout the film, which unambiguously demonstrates true global alignment from solution-based SWCNTs.
Finally, based on experiments tuning the electrostatic environment, we propose that charge ordering on the filter membrane is one of the driving forces involved in the spontaneous alignment of SWCNTs along a common axis. 

\begin{figure*}
	\centering
	\includegraphics[width=0.97\textwidth]{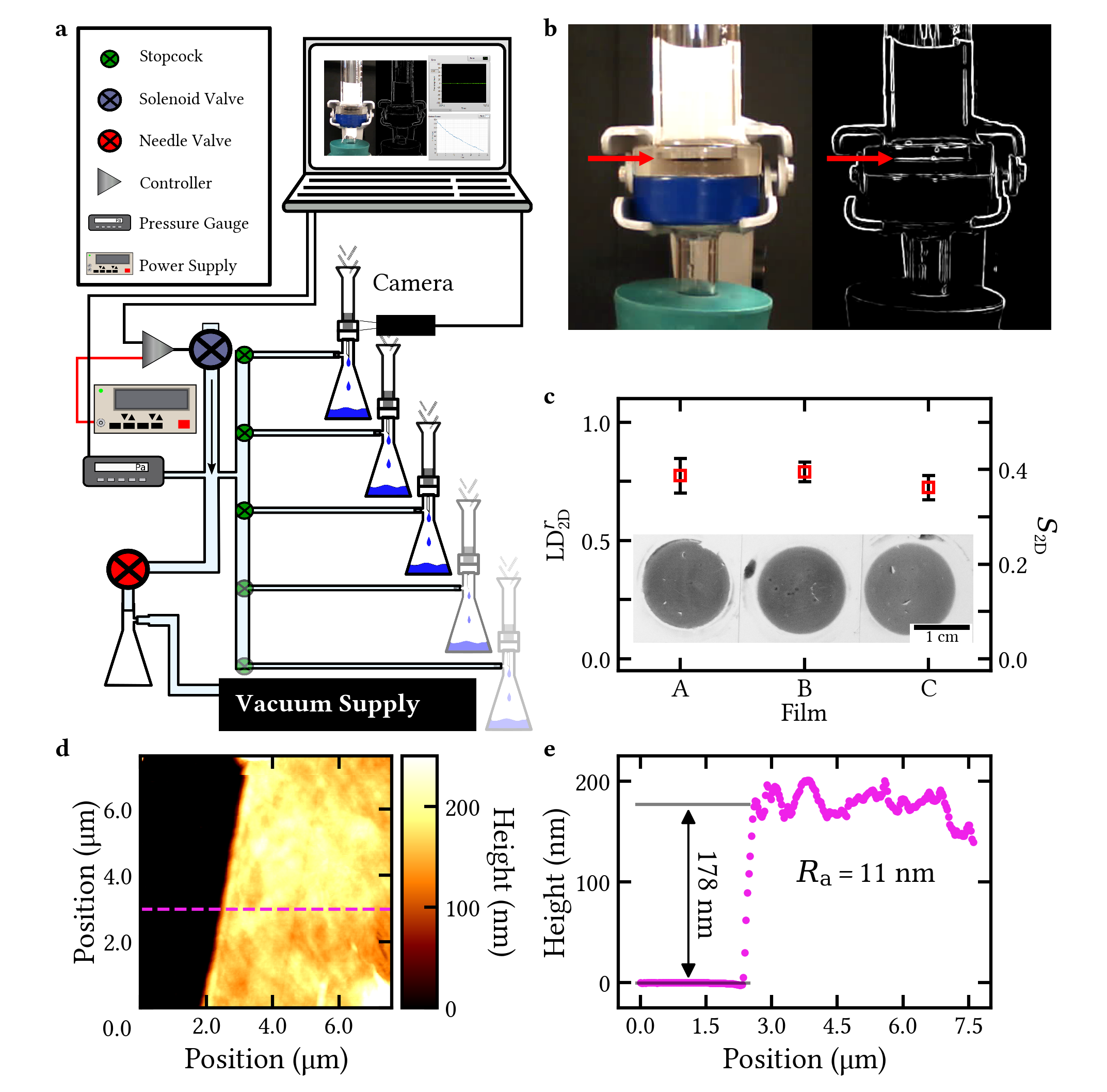}
	\caption{(a) Schematic of the automated, parallelized filtration assembly. The ability to simultaneously create multiple aligned SWCNT films using machine vision and a feedback-control system enables reproducible and scalable production.  
    (b) (left) Image of the filtration assembly taken by the camera during filtration.  (right) Corresponding contrast-enhanced edge image produced using the Canny edge algorithm. The arrows in both figures highlight the meniscus edge, which we detect, numerically fit, and monitor during filtration.  
    (c) Two-dimensional reduced linear dichroism, \ldr, for three SWCNT films made in parallel.  The high and uniform degree of alignment confirms that each separate arm of the parallelized system is equivalent.  Inset: pictures of the corresponding aligned SWCNT films. (d) Atomic force microscopy (AFM) height map of film made with 0.030 wt.\%~DOC.
    (e) AFM height profile extracted from the magenta line on (d). Measured film thickness of 178$\pm$11~nm is on the order of one nanotube length. From the AFM film profile, a surface roughness, $R_{\rm a}$, of 11~nm is obtained.
\label{fig:setup}}
\end{figure*}

To prepare our individualized SWCNT solution for filtration, we use Carbon Solutions (Riverside, CA) P2 arc-discharge SWCNTs (lot\# 02-A011), which are dispersed via tip sonication~\cite{NISTDisclaimer} in 20~g/L sodium deoxycholate (DOC) in H$_2$O and then centrifuged to remove non-SWCNT contaminants.  
To mitigate discrepancies between water- and non-water-filled SWCNTs, we used nanotubes filled with C$_7$H$_{16}$~\cite{CampoNanoHoriz2016}.
Next, a rate-zonal centrifugation method is applied to sort the nanotube solution to remove bent and very short SWCNTs.
At the end of this multi-step procedure~\cite{FaganACSNano2011}, which also includes removal of solution components other than H$_2$O, DOC, and SWCNTs via ultrafiltration and increasing the SWCNT concentration, we have a mixed-chirality SWCNT solution composed of long, straight tubes, in 10~g/L DOC, which are ideal for this alignment technique.
Before filtration, the SWCNT dispersion is diluted to have a DOC concentration of 0.03~wt.\% and an approximate SWCNT concentration of 8~$\mu$g/mL, which was determined through optical density measurements.

To automate the filtration setup shown in Figure~\ref{fig:setup}a, we developed software to detect, numerically fit, and control the edge of the SWCNT solution meniscus. 
Meniscus detection and tracking was performed using an adaptable detection algorithm to convert the real image to an edge outline, as shown in Figure~\ref{fig:setup}b.  We developed software to numerically fit the meniscus edge from this outline, which significantly reduced fluctuation-created false positives.  
Because the meniscus is tracked as a function of time, it serves as a measure of flow rate.
Pressure is maintained via a PID-controlled variable leak using a proportioning solenoid valve with an applied source vacuum of 28.8~kPa.
Accurate and precise ($\pm$2~Pa) applied pressure control is achieved across a broad pressure range of over four orders of magnitude from 3~Pa to 6000~Pa.
The pressure is increased at the end of the filtration process to dry the film and prevent disrupting the still-wet SWCNT structure. 

\begin{figure*}
	\centering
	\includegraphics[width=0.97\textwidth]{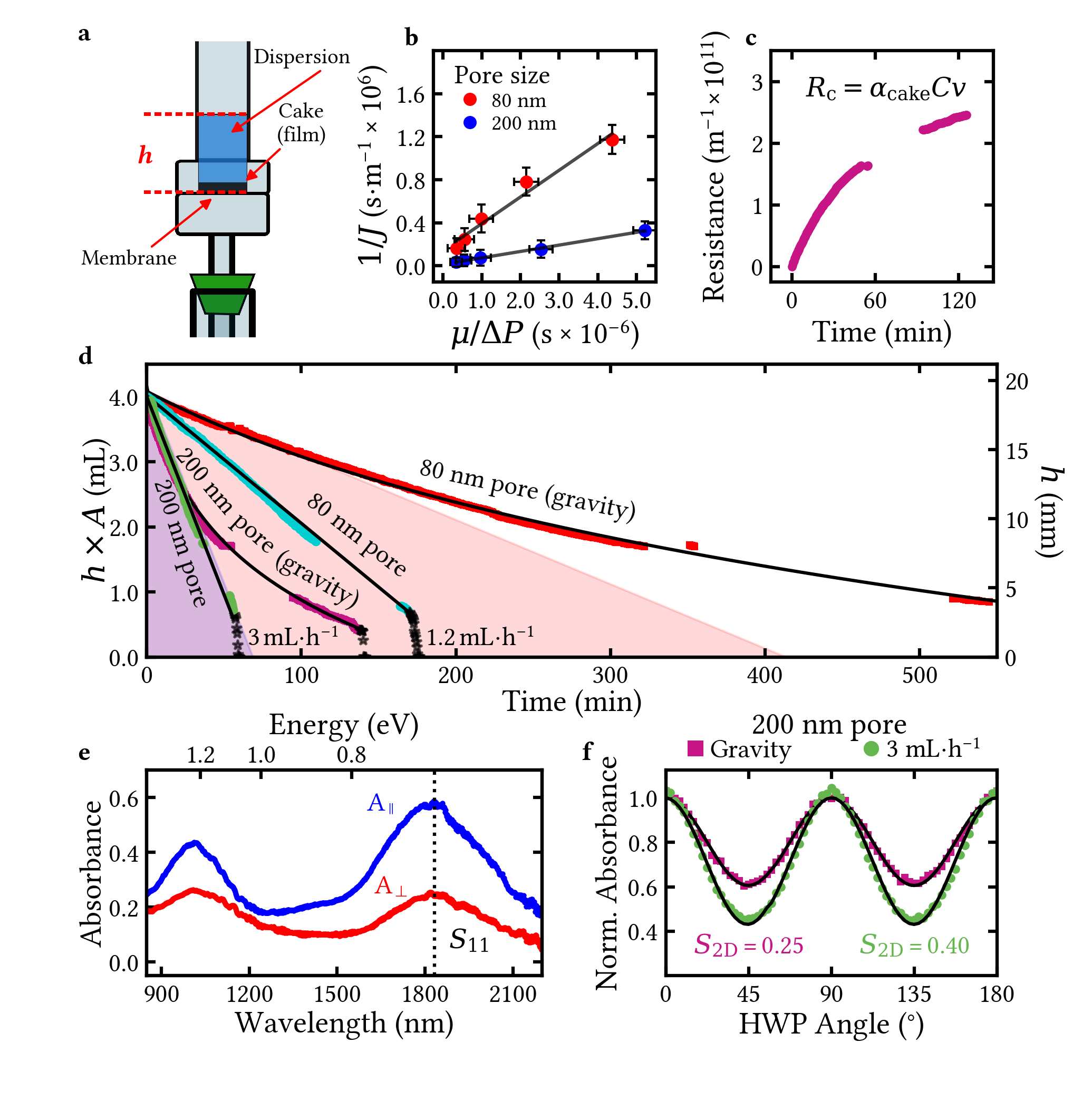}
	    \caption{
	    (a) Graphical depiction of the filtration setup.  The height of the SWCNT dispersion, $h(t)$, is measured as a function of time, $t$, by a computer-monitored camera. 
    (b) Reciprocal of the permeation flux, $J$, as a function of $\mu/\Delta P$ for the 80~nm (red) and 200~nm (blue) pore size membranes. The measured membrane resistance, $R_\mathrm{m}$, is equal to $2.5\times 10^{11}$~m$^{-1}$ for the 80~nm pore size membrane and $3.8\times 10^{10}$~m$^{-1}$ for the 200~nm one.
    (c) Cake (i.e., the SWCNT film during filtration) resistance, $R_{\rm c}$, as a function of time. As more SWCNTs are deposited on the membrane surface, the overall resistance to flow increases. Data shown for a 80~nm pore size membrane and a SWCNT concentration of 8~$\mu$g/mL.
    (d) Time-dependence of the measured SWCNT dispersion volume for different filter pore sizes and pressure conditions. The gravity filtration data and fit (black curve) denote the flow rate lower limit that is possible with a given membrane pore size. The shaded regions for both the 80~nm (pink) and 200~nm (purple) membrane pore sizes indicate the accessible region of achievable flow rates.  Using our results from (b) and (c), we produce a constant flow rate for the 80~nm and 200~nm filter membranes.  Gaps in the data are created when the meniscus goes behind the glass funnel lip.  Starred points denote the film drying process, where the pressure is increased to remove residual water.  
    (e) Polarized absorbance spectra of a SWCNT film produced using the 200 nm pore membrane. A half-wave plate (HWP) rotates the light polarization with respect to the alignment direction.  The black dashed line indicates the $S_{11}$ absorbance feature of one of the nanotube chiralities.   
    (f) Normalized polarized $S_{11}$ absorption of a film made with the 200 nm pore membrane using a constant flow rate (green) and gravity filtration (purple).  
\label{fig:flowrate}}
\end{figure*}

Besides eliminating human control from the alignment process, our machine vision-based SWCNT film configuration easily allows for production upscaling. 
As detailed in Figure~\ref{fig:setup}a, although just one filtration rig (`master') is monitored, multiple films are simultaneously produced by putting filtration assemblies in parallel.  
Since each parallelized assembly is nearly identical to the master rig, excellent alignment for several SWCNT films is achieved.
As an example of this high degree of uniformity achieved by our system, we show in Figure~\ref{fig:setup}c three films simultaneously produced in our parallelized system, which have nearly identical 2D nematic ordering ($S_{\rm 2D}$).  
We use $S_{\rm 2D}$ as a measure of SWCNT alignment in our films throughout this work (measurement details described below, in Figures S1 and S2, and the corresponding Supplementary Information text).  
As seen in Figures~\ref{fig:setup}d and e, the SWCNT films are 178$\pm$11~nm thick (in this paper, uncertainty is reported as $k = 1$ standard deviation), which is on the order of the length of one nanotube ($\approx$200~nm to 400~nm). Nanotube film thicknesses on this order (and smaller) are often treated as 2D constructs, because of the dimensionality of their physical properties~\cite{WangPhysRevMater2018}.  We note that the film thicknesses measured here are on the high end of the thickness spectrum compared to those reported by prior groups~\cite{HeNatureNano2016, ChiuNanoLett2017}, thus supporting our claim of excellent global alignment over a substantial SWCNT deposition amount. 

The combination of automation and parallelization enables us to produce multiple copies of aligned films under a wide range of different physical and chemical conditions.  
Previous SWCNT alignment protocols have been unable to achieve a constant filtration flow rate, which results in a time-varying SWCNT cake (i.e., film during filtration; Figure~\ref{fig:flowrate}a) deposition rate, thus hindering optimal SWCNT alignment.  
In order to achieve a constant filtration flow rate, which is expressed as a permeation flux, $J$ ($= \frac{\rm flow~rate}{\rm area}$), we first had to empirically determine the membrane resistance, $R_{\rm m}$, and cake resistance, $R_{\rm c}$:

\begin{equation}
\frac{1}{J}=\frac{\mu (R_{\rm m}+R_{\rm c})}{\Delta P},                                                                       
\end{equation}

\noindent where $\mu$ is the viscosity of the permeate (SWCNT solution), $\Delta P$ is transmembrane pressure, and $R_{\rm c} = \alpha_{\rm cake} C v$ with $\alpha_{\rm cake}$ as the specific cake resistance, $C$ the dispersion concentration, and $v$ the filtrate volume per unit area (see Figure S3 for information about the measurement of $\alpha_{\rm cake}$).  
$R_{\rm m}$, which is independently measured through a controlled water filtration experiment shown in Figure~\ref{fig:flowrate}b, is 2.5$\times$10$^{11}$~m$^{-1}$ for the 80~nm pore-size membrane and 3.8$\times$10$^{10}$~m$^{-1}$ for the 200~nm pore-size membrane. 
While $R_{\rm m}$ is nearly constant throughout the filtration process, $R_{\rm c}$ increases with time as the cake is deposited, which is shown in Figure~\ref{fig:flowrate}c.  
Using our determination of the time-independent $R_{\rm m}$ and the time-dependent $R_{\rm c}$, we can then tune $\Delta P$ throughout the film deposition to keep $J$ constant.  
It should be noted that $\Delta P$ is the total (time-dependent) transmembrane pressure, which is the sum of the applied pressure, $P_{\rm applied}$, and the head pressure, $\rho g h(t)$, where $\rho$ is the dispersion mass density, $g$ is gravitational acceleration, and $h(t)$ is the time-dependent solution column height.

Figure~\ref{fig:flowrate}d shows our ability to achieve a constant flow rate for two membrane pore sizes (80~nm, cyan trace; and 200~nm, green trace) throughout the entire film deposition process.
The gravity-driven filtration (i.e., no applied pressure; red and purple traces) curves demarcate the slowest flow rate possible in our system with a specific $R_{\rm m}$, while the colored regions (pink and purple for 80~nm and 200~nm pore-size membranes, respectively) indicate the possible range of flows accessible when external pressure is applied (see Figure S4 for a more-detailed view of Figure~2d).  
The high value of $R_{\rm m}$ for the 80~nm pore-size membrane allows for a greater variability of the flow rate, but often at lower values of $J$.  
Critically, the procedure that we delineate here, measuring $R_{\rm m}$ and $R_{\rm c}$ and then tuning $\Delta P$ to maintain a constant $J$, means that we can apply the parameters described in this work (flow rate, DOC concentration, SWCNT concentration, etc.) to different SWCNT types (laser oven, HiPCO, CoMoCAT, arc discharge, etc.) and membranes with differing pore sizes and materials.  
As such, our procedure greatly broadens the applicability and utility of this filtration-based alignment method.

Throughout this Letter, we use several optical spectroscopic techniques to determine $S_{\rm 2D}$ for our SWCNT films after they have been transferred to either a glass coverslip or quartz substrate (see Figure S5 for details regarding the film transfer).  
Since SWCNTs have a highly anisotropic absorption coefficient, $\alpha$, polarized optical spectroscopic techniques are commonly used to measure nematicity in SWCNTs.  
Although these methods are not without their limitations, they provide a straightforward way to measure $S_{\rm 2D}$.  
In this work, we rely on three polarized optical methods to determine nematicity:  reduced linear dichroism, \ldr, polarized Raman scattering (Figure S1), and birefringence ratios (Figure S2), whose precise formulations are given in the Supplementary Information.  
The guiding principle behind all three methods, however, is graphically captured by Figure~\ref{fig:flowrate}e, which shows that when the optical electric field is parallel to the SWCNT axis, the on-axis absorption coefficient, $\alpha_{\parallel}$, is high; in contrast, when the electric field is orthogonal to the SWCNT axis, the off-axis absorption coefficient, $\alpha_{\perp}$, is suppressed.  Depending on the optical technique, we either rotate the light polarization angle using a half-wave plate (HWP), or keep the light polarization fixed and rotate the SWCNT film; in both cases, $\alpha_{\parallel}$ and $\alpha_{\perp}$ are probed.

Figure~\ref{fig:flowrate}f compares the LD$^{r}_{\rm 2D}$ $[=\frac{2(\alpha_{\parallel}-\alpha_{\perp})}{\alpha_{\parallel}+\alpha_{\perp}} =$ $\frac{2(A_{\parallel}-A_{\perp})}{A_{\parallel}+A_{\perp}}]$ of films made via gravity-driven filtration (purple) and using a constant flow rate filtration (green).  
Although the $S_{11}$ absorption used to calculate \ldr is composed of both SWCNT and non-SWCNT components that are not related to excitonic transitions (thus, underestimating the true value of $S_{\rm 2D}$), it is very clear from this figure that greater alignment (i.e., a higher \ldr value) is achieved when the flow rate is fixed.  
Thus, in addition to improving reproducibility and allowing us to generalize our parameters to other SWCNT types, precise control of the pressure to make the flow rate constant also greatly enhances SWCNT nematicity.  


\begin{figure*}
	\centering
	\includegraphics[width=0.97\textwidth]{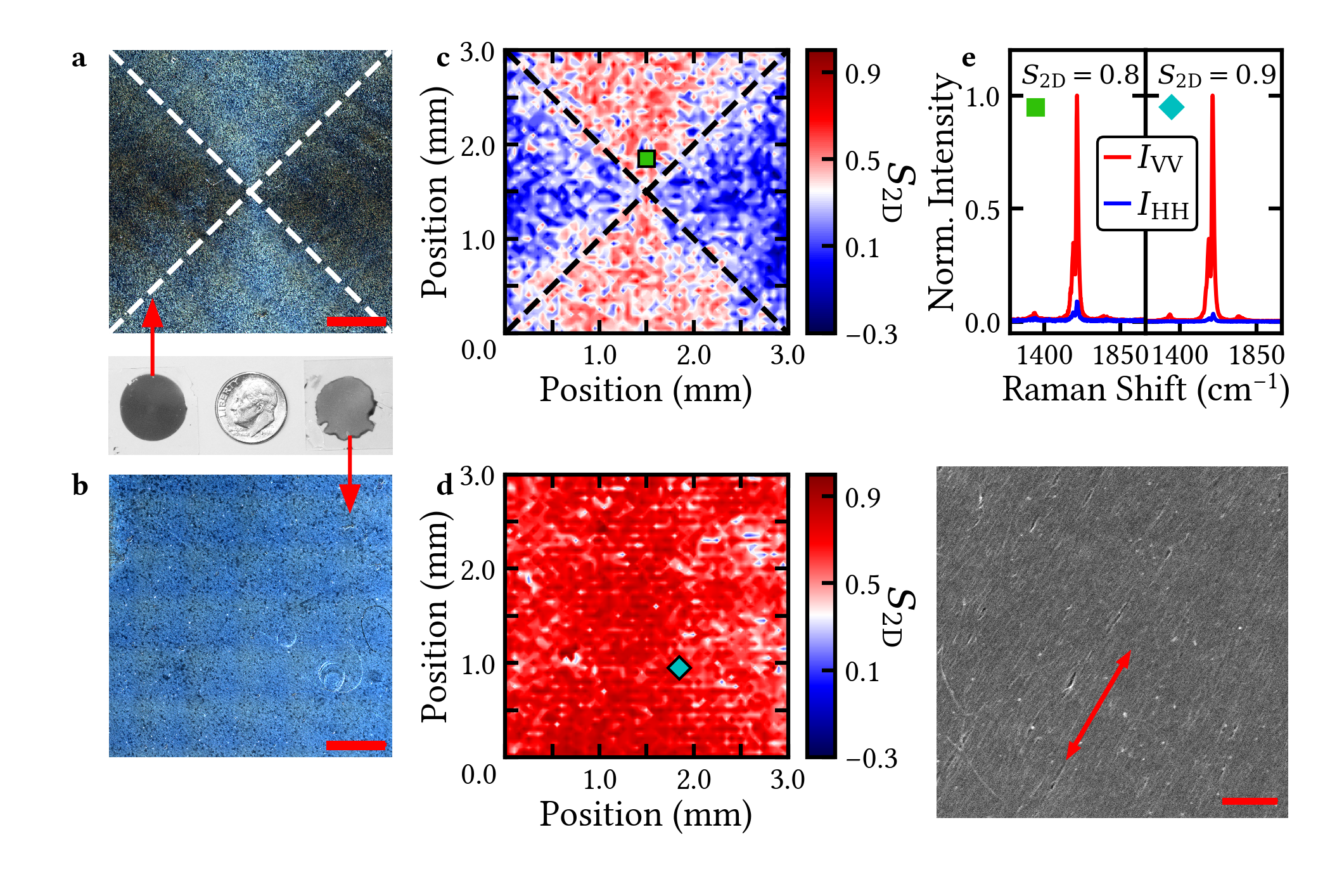}
	\caption{(a)  Cross-polarized microscope image showing the formation of a 2D spherulite (radial surface SWCNT alignment due to meniscus combing) on the front side of the SWCNT film. Scale bar: 1500~$\mu$m. 
   (b) Cross-polarized microscope image of a SWCNT film created using silanated glassware, a procedure which significantly flattens the meniscus.  Despite imaging over a large area of the film, there are no signs of a 2D spherulite. Scale bar: 1500~$\mu$m.
   (c) Radial alignment of the SWCNTs (and uneven cake deposition) is created by meniscus combing at the end of the filtration process.  The red arrows indicate the radially aligned force director at the cake surface.
    (d) Spatial map of $S_{\rm 2D}$ of the front side of the film in (a) as measured by polarized Raman spectroscopy at 532 nm.  The formation a radial pattern and a lack of global alignment across the front surface of the SWCNT film is clearly observed. The square marker indicates where the Raman spectrum shown in (f) is taken.
    (e) Spatial map of $S_{\rm 2D}$ for the front side of the film in (b).  The high degree of SWCNT alignment over a large area and distinct lack of a 2D spherulite formation provide strong evidence that the meniscus flattening eliminates SWCNT radial polarization. The diamond marker indicates where the Raman spectrum shown (f) is taken.
    (f) Polarized Raman spectra of films in (a) (left) and (b) (right). The red traces ($I_{\rm VV}$) are measured with the incident and scattered light both polarized along the SWCNT film axis, while the blue traces ($I_{\rm HH}$) are measured with the incident and scattered light polarized perpendicular to this axis.  These measurements are necessary for calculating $S_{\rm 2D}$.  
    (g) Scanning electron microscopy image of a SWCNT film made with 0.030 wt.\%~DOC. Arrow indicates alignment axis. Scale bar: 500~nm.
\label{fig:spherulite}}
\end{figure*}

Along with flow rate control, we also detect and address a previously unreported effect in SWCNT films created by the meniscus of the SWCNT dispersion during filtration.  Specifically, our use of spatially-resolved polarized optical techniques, such as birefringence and polarized Raman scattering mapping, reveals the presence of a \textit{radial} SWCNT alignment on the front surface of our films (i.e., the side that faces upward during filtration).  
This type of alignment (at least in three dimensions) is known as a spherulite and is commonly observed in films of 1D crystals when the solution meniscus combs (or drags) across the film surface, which produces a force director that radially polarizes the crystals (Figure~\ref{fig:spherulite}c). 
As Figures~\ref{fig:spherulite}a and d clearly indicate, we observe this radial alignment on the surface of our SWCNT films due to meniscus combing during the final stages of our filtration; we refer to this feature as a 2D spherulite.  
To remove this meniscus-created radial orientation, we increased the hydrophobicity of our filtration glassware using a silanation procedure (see Figure S6 and the associated text in the Supplementary Information for details).
This technique greatly flattens the meniscus and prevents it from dragging across the SWCNT cake at the end of filtration (Figure~\ref{fig:spherulite}c).
Figures~\ref{fig:spherulite}b and e show the results of flat-meniscus filtration.  
In stark contrast to the traditional SWCNT alignment method, the nanotubes on \textit{both} sides of the film are well ordered (see Figures S7 and S8 for Raman maps of both front and back film surfaces).  
Importantly, this double-sided alignment extends across large distances, which demonstrates global SWCNT nematic ordering.

Interestingly, depending on the thickness of the film, the skin depth of the optical probe, and whether the probe measures reflection or transmission, the use of polarized optical spectroscopy may not unambiguously detect the presence of a 2D spherulite. 
This point is strengthened when one remembers that after the film is transferred from the membrane to a substrate, the front film surface during filtration is now the back film surface for optical measurements, which may not always be as well measured as the front surface.  
Figures~\ref{fig:spherulite}f and S7c capture the non-obvious signature of the 2D spherulite behavior:  although $S_{\rm 2D}$ is clearly increased when the 2D spherulite is removed through silanating the glassware, the effect on the measured nematicity is not huge.  
Only through the careful spatial mapping of \textit{both} SWCNT film surfaces does the radial polarization, and thus lack of global ordering, appear. 
The high degree of alignment across the non-spherulite film is additionally supported through direct imaging techniques like scanning electron microscopy (Figure~\ref{fig:spherulite}g), which clearly indicates excellent SWCNT alignment.

\begin{figure*}
	\centering
	\includegraphics[width=0.97\textwidth]{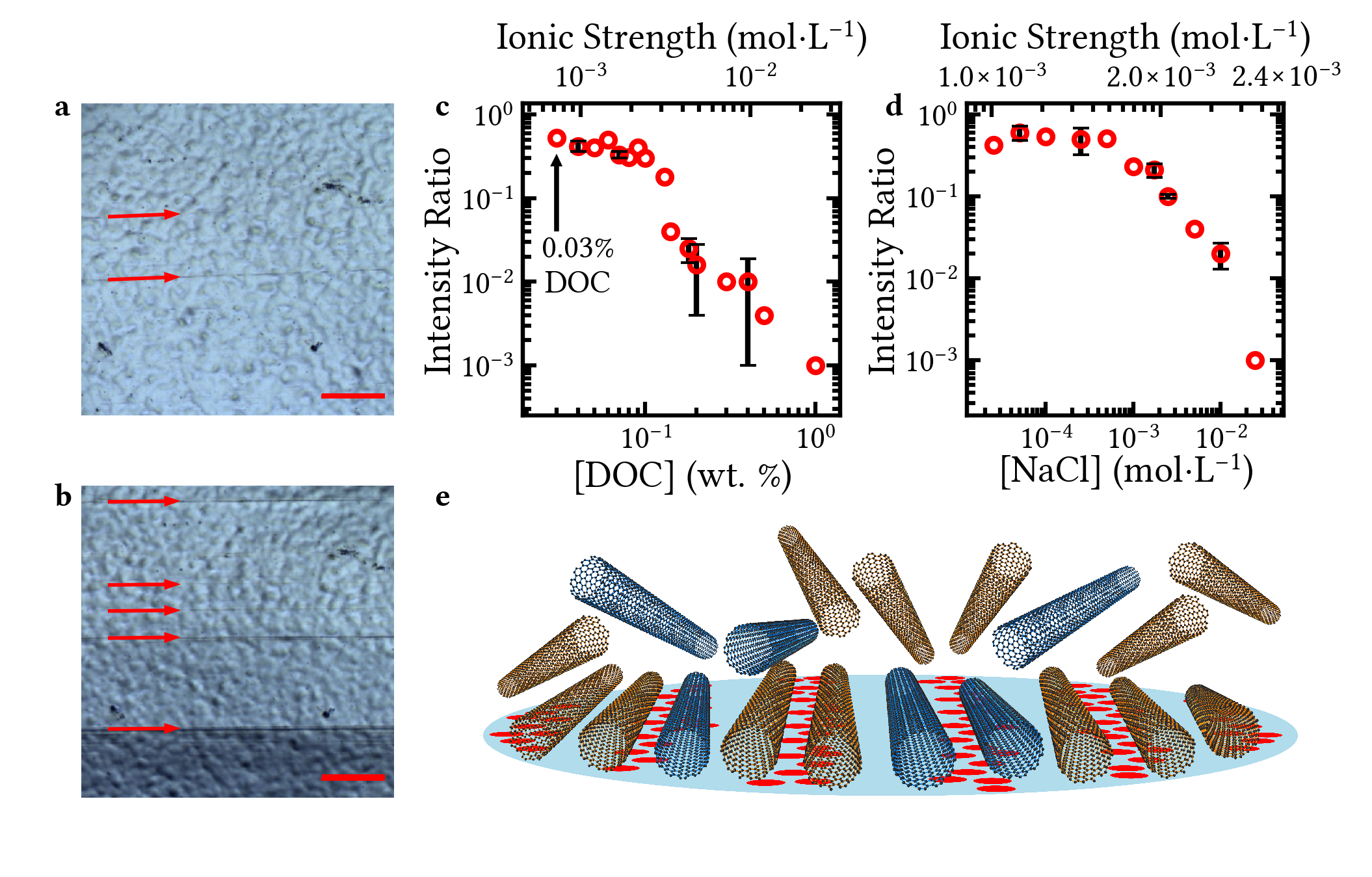}
	    \caption {(a) Microscope image of a membrane used for the filtration process.
    Small grooves (indicated by the red arrows) are observed on the surface, which are created from the membrane production process. Scale bar: 250~$\mu$m.
    (b) Although groove density is increased when the membrane PVP surface in (a) is swept with ethanol-wetted wipe, we find that the grooving is not as important as tribocharging. Scale bar: 250~$\mu$m.
    As the concentration of (c) DOC and (d) NaCl increases (i.e., inter-SWCNT electrostatic interactions decrease), the birefringence intensity ratio of the SWCNT film is sharply depressed.  These behaviors confirm that the inter-nanotube electrostatic environment is critical to the formation of aligned SWCNT films. For clarity, representative error bars are shown for five points.
    (e) Graphical depiction of our proposed model for ordered SWCNT formation.  Charges puddle on the filtration membrane in quasi-linear chains created by intentional wiping.  These electrostatic fields, in competition with the complex electrostatic inter-nanotube environment, compete to produce an aligned SWCNT phase on the membrane.
\label{fig:electrostatic}}
\end{figure*}

We also investigated the SWCNT alignment mechanism using our automated filtration system with the previously described improvements to the alignment protocol.  
Figure~\ref{fig:electrostatic}a shows grooves in the filter membrane created during the manufacturing process.  
Although these grooves are not uniformly spaced or appropriately sized for the nanotubes we are using, they still raise an important question about whether or not they play a role in nanotube alignment.  
To help address what part, if any, these channels play in nanotube alignment, we augment the grooving by sweeping an ethanol-wetted wipe across the top surface of the filter membrane along the initial groove direction (Figure~\ref{fig:electrostatic}b) prior to filtration.  
As shown in Figure S9, the ethanol-wiped filter membrane increases the SWCNT film alignment considerably going from a $S_{\rm 2D}$ of 0.26 to 0.52.
To distinguish between whether the increased membrane grooving or electrostatic charging (or both) are primarily responsible for the enhanced SWCNT nematicity, we swept a filter membrane with a wipe wetted with water and another with ethanol (Figure S9) and showed that $S_{\rm 2D}$ increased from 0.27 for the SWCNT film made from water-wiped membrane to 0.42 with the ethanol-wiped membrane.
This enhancement strongly suggests that ordered charging, instead of mechanical grooving, is responsible for increased SWCNT alignment.
We note that a similar increase in SWCNT alignment was observed in self-assembled SWCNT nanowires when the glass substrates were directionally wiped with ethanol~\cite{HobbieACSNano2009}.

Along with charging the membrane, inter-SWCNT electrostatic interactions are the other major factor determining $S_{\rm 2D}$.  As shown previously~\cite{HeNatureNano2016}, the DOC concentration of the SWCNT solution strongly impacts the achievable $S_{\rm 2D}$ (Figure~\ref{fig:electrostatic}c), because the Debye interaction length between SWCNTs decreases with increasing DOC coverage of the nanotube.  The measured $S_{\rm 2D}$ scaling with DOC concentration strongly guided our decision to use solutions with 0.030 wt.\% DOC concentration, which is over an order of magnitude below the critical micelle concentration~\cite{EspositoJPhysChem1987, DalagniLangmuir1997}.  Further experiments using NaCl to tune the ionic strength confirm that as the tube-tube electrostatic interactions decrease due to increased screening, the SWCNT nematicity plummets (Figure~\ref{fig:electrostatic}d).

The strong role played by directional tribocharging and inter-tube electrostatics suggest that linear arrays of charges are accumulating on the membrane, as depicted in Figure~\ref{fig:electrostatic}e.  
Although mechanical membranes grooves may play a minor role in nanotube alignment, the large size of the grooves, lack of groove uniformity, and their relative irregularity all suggest that they are not directly responsible for SWCNT alignment.  
Instead, we propose that the wiped filter membranes acquire some small net charge that is linearly arranged, which in combination with tube-tube interactions, creates alignment along a common axis. 
Given the estimated net charge magnitude on the filter membrane, it is unsurprising that this charge-directed alignment is quite easy to destroy, as seen by small changes in the ionic strength of the SWCNT solution.
Directional charging has been observed in a host of situations by numerous researchers, including efforts to pattern surface charges~\cite{BurgoLangmuir2012, ZhuNanoLett2013} and to align particles via electrostatic puddling~\cite{GrzybowskiNatureMater2003}. Further work on how to enhance and control charge arraying on filter membranes is ongoing.

In summary, we have created an automated, parallelized SWCNT filtration system that can create simultaneous and reproducible SWCNT films with a high degree of true global alignment. 
We find that holding the filtration flow rate constant using our pressure-controlled system enhances the nematic order of our films. 
In addition, we both measure and remove 2D spherulite formation on the front-surface of the SWCNT films by flattening the meniscus using silanated glassware.  
We propose that directional charging on the filter membrane and inter-SWCNT electrostatic interactions are the two driving forces behind the alignment of nanotubes using this filtration technique.  
Our innovations on the SWCNT filtration method, as well as the results described here, pave a clear path for both research- and industrial-scale implementation of highly aligned SWCNT films from aqueous solutions.

\begin{acknowledgement}

The authors thank Dr.~Joseph R.~Murphy for his input on the machine-vision implementation and AFM data processing, and Dr.~John Ackerman for his help with the glassware silanation.  TAS acknowledges funding from the W.~M.~Keck Foundation, and JSW and WDR acknowledge funding from the School of Energy Resources at the Univ.~of~WY.

\end{acknowledgement}

\newpage 

\renewcommand{\thepage}{S\arabic{page}}  
\setcounter{page}{1}
\renewcommand{\thesection}{S\arabic{section}}   
\renewcommand{\thetable}{S\arabic{table}}   
\renewcommand{\thefigure}{S\arabic{figure}}
\setcounter{figure}{0} 

\onecolumn

\noindent \textbf{Supplemental information}

\section{Preparation of single-wall carbon nanotube (SWCNT) \\dispersions}
Alkane-filled SWCNT dispersions were generated in the manner of Campo et al.~\cite{CampoNanoHoriz2016}.  
Briefly, electric arc-synthesized SWCNTs (Carbon Solutions, Riverside CA, P2 grade, lot\# 02-A011) were incubated in neat n-heptane (EMD Millipore) for $>$12~hours, which filled the entire nanotube population with the alkane.
After incubation, the \ce{C7H16}@SWCNT powder was filtered against a membrane (VVLP, 0.1~$\mu$m, Millipore) and allowed to fully dry at room temperature. 

The \ce{C7H16}@SWCNT soot was dispersed in multiple $\approx$40~mL aliquots via tip sonication (45~minute, $\approx$0.9~W/mL), in each case with the vial in an ice-water bath at a nominal concentration of 1~mg/mL of SWCNTs in 20~g/L sodium deoxycholate (DOC, Sigma BioXtra) in water solution.
Initial purification on the combined aliquots was performed via centrifugation in a J2-21 high-speed centrifuge (JA-20 rotor, 1885~rad/s (18~kRPM), 2~hours), after which the resulting supernatant was collected.
This sonicated-centrifuged dispersion was subsequently layered ($\approx$8.2~mL) above a dense race layer (28~mL) comprised of 10~wt./v\% iodixanol (Sigma, sold as Optiprep) and 10~g/L DOC for rate-zonal purification (VTi 50 rotor, 5236~rad/s (50~kRPM), 2~h 45~minute, 20$^{\circ}$C) in a Beckman L80-XP ultracentrifuge collecting the main band in the center of the tube after centrifugation.
Stirred ultrafiltration cells (Millipore) were used to both reduce the iodixanol concentration to $\ll$1~$\mu$g/mL and to concentrate the SWCNT dispersion to $\geq$1~mg/mL in 10~g/L DOC solution as determined by absorbance spectroscopy using an extinction coefficient of 2.1~A\! mg/mL\! mm at 850~nm.

\newpage

\section{Details of the filtration system}
The filtration assembly consisted of three main components: the borosilicate funnel (15~mL), the stainless steel (SS) mesh frit (25~mm), and the borosilicate glass frit support.
The assembly also used a spring-loaded clamp for securing the funnel in place, a silicone stopper, and two PTFE gaskets, one of which is used under the SS frit, and the other placed on top of the wetted, hydrophilic membrane.

Once the filtration system is assembled and connected to a B\"{u}chner flask for vacuum filtration, a vacuum source is attached and a 28.8~kPa pressure is applied.
The source vacuum is then regulated through the use of a needle valve and proportioning solenoid valve (PSV), which acts as a controllable leak.
The PSV (normally closed) runs on a 16~V-source voltage, which is controlled using a separate 5~V-modulated controller via a computer-controlled digital-to-analog converter.
The PSV and a digital pressure gauge are used to control the pressure of the system.
Through the use of stopcock valves, we are then able to implement parallel assemblies that can be turned on or off as desired. 

Various pore sizes of wetted, hydrophilic polyvinylpyrrolidone (PVP)-coated membranes~\cite{HeNatureNano2016} are then used for carrying out the experiments described in the main text.
4~mL of a SWCNT dispersion is carefully pipetted into the assembly funnel, so as to not mechanically disturb the individualization of SWCNTs~\cite{FernandesJColloidInterSci2017}.
The desired pressure(s) to be applied for controlled flow rate are then loaded into the software program, and the program is started.
The region of interest in the camera image of the filtration assembly is selected such that the triggering event will occur when the remaining volume is near 0.7~mL.
When the trigger event occurs, the pressure is increased such that the flow rate falls between 10~mL/h to 15~mL/h~\cite{HeNatureNano2016}; this accelerated pressure is indicated with stars in Figure 2d of the main text and Figures~\ref{fig:flowrate}a and b.

To implement machine-vision detection of the meniscus, a few modifications were made to the assembly to allow the software to easily and readily identify edges and discontinuities.
These changes to the system included: masking the back of the filtration funnel with white vinyl tape, masking the frit support with blue vinyl tape, and placing the assembly in front of a black backdrop.
The blue, white, and black colors create contrast, which clearly defines the edges for detection and tracking.
The white tape masking the back of the funnel was chosen to provide contrast between different SWCNT dispersions and the funnel itself, the blue tape defines the bottom edge of the funnel, and the black backdrop defines the outline of the funnel.
These differences are seen in Figure 1b of the main text. 

\newpage

\section{Polarization-sensitive optical characterization}
For optical characterization a number of different polarized methods were used including: linear dichroism, spatially-resolved Raman spectroscopy, and birefringence microscopy.
\subsection{Polarized absorbance}

In the case of polarized absorption, we utilize the anisotropic absorbance of SWCNTs to quantify \stwod. 
Since the absorption coefficient of a $J=\pm 1$~$(V_i \leftrightarrow C_i)$ excitonic transition, $\alpha$, is fully realized only when the incident optical field is parallel to the SWCNT, we can define two absorption coefficients, $\alpha_\parallel$ and $\alpha_\perp$, corresponding to the two physical extremes of the incident field parallel and perpendicular, respectively, to the SWCNT axis.
In the main text, we used 2D reduced linear dichroism, \ldr, to estimate the two-dimensional nematic order parameter, \stwod. Following Katsutani et al.\cite{KatsutaniPRB2019}, we define \ldr\ as:
\begin{equation}
\mathrm{LD}^{r}_{\mathrm{2D}} = \frac{2(\alpha_{\parallel} - \alpha_{\perp})}{\alpha_{\parallel}+\alpha_{\perp}},
\label{eq:ldr}
\end{equation}
\noindent where
\begin{equation}
S_{\mathrm{2D}} = \frac{\mathrm{LD}^{r}_{\mathrm{2D}}}{2} = \frac{(\alpha_{\parallel} - \alpha_{\perp})}{\alpha_{\parallel}+\alpha_{\perp}}.
\label{eq:StwoD}
\end{equation}

\subsection{Spatially-resolved polarized Raman spectroscopy}
In measuring spatially-resolved polarized Raman spectroscopy, a five-axis stage is used:  three of the axes are controlled via linear actuators, while the other two axes are used to adjust pitch and yaw to correct for sample tilt.
For spatial mapping, we take points in 50~$\mu$m steps over a 3~mm $\times$ 3~mm area.  A Mitutoyo, long-working distance, 50x objective with an estimated spot size of $\approx$1~$\mu$m is used to focus and collect Raman scattering from a 532 nm continuous-wave excitation source.  Spectra were resolved on a 750 nm blazed, 1200 grooves/mm grating using a 320 mm Isoplane spectrometer and a LN$_2$-cooled Si CCD (all from Princeton Instruments).

To determine \stwod\ using polarized Raman spectroscopy, we use the following equation:~\cite{Zamora-LedezmaNanoLett2008}

\begin{equation}
S_{\mathrm{2D}} = \frac{\Delta I_{\mathrm{VV}} - I_{\mathrm{HH}}}{\Delta I_{\mathrm{VV}} + (1 + \Delta)I_{\mathrm{VH}} + I_{\mathrm{HH}}}.
\label{eq:sRaman}
\end{equation}
Here, $I_{\mathrm{VV}}$, $I_{\mathrm{VH}}$, $I_{\mathrm{HH}}$ are taken to be the integrated intensity of the G band, and $\Delta$ is the dichroic ratio $\Delta=\frac{A_{\parallel}}{A_{\perp}}$, while VV, VH, and HH define the orientation of the incident and analyzed polarization with respect to the SWCNT alignment axis, respectively. The orientations are either parallel, parallel (VV), parallel, perpendicular (VH), or perpendicular, perpendicular (HH).
\begin{figure*} [h]
	\centering
	\includegraphics[width=0.75\textwidth]{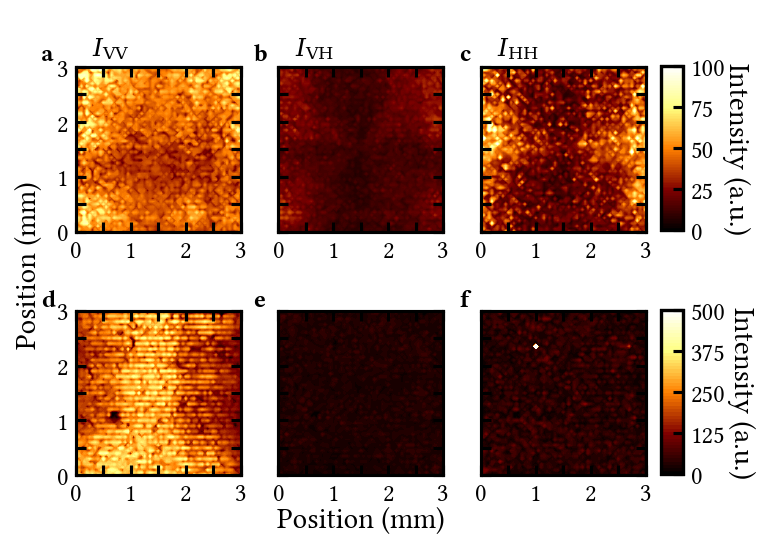}
	\caption{
	(a), (b), and (c) show the spatially-resolved polarized Raman spectroscopy maps of $\ivv$, $\ivh$, and $\ihh$, respectively, used in the calculation of the \stwod\ nematic order parameter in Figure 3d of the main text. The same Maltese-cross pattern that is apparent in the main text is also visible here, primarily in (b). 
	(d), (e), and (f) show the $\ivv$, $\ivh$, and $\ihh$, respectively, maps used in the calculation of the spatially-resolved \stwod~map in Figure 3e of the main text. It can be seen here that the alignment is highly uniform over a 9~mm$^{2}$ area. 
\label{fig:ramanMaps}}
\end{figure*}

When carrying out this measurement, the stage is moved to an initial starting point, (0,0), where a polarized absorption measurement is taken.
This measurement consists of rotating a half-wave plate (HWP), thus changing the orientation of the incident linear polarization to determine the SWCNT alignment axis, as well as to measure $\Delta$.
The two angles of the HWP that yield $A_{\parallel}$ and $A_{\perp}$ are then used as the positions that define the incident polarization orientations for scanning over the sample.
Before starting the 2D scan over the sample, the incident power is measured at the sample for both angular positions and is then held constant for the duration of the measurement.
Results from this method can be seen in \autoref{fig:ramanMaps} where the individual 2D maps ($I_{\rm VV}$, $I_{\rm VH}$, and $I_{\rm HH}$) show the integrated intensity of the G-band for the calculation of \stwod for Figures 3d and e in the main text.
Interestingly, the radial polarization causing the formation of the 2D spherulite can be distinguished in the $\ivh$ map of the SWCNT film. 

\subsection{Polarized microscopy and the birefringence intensity ratio}

\begin{figure*}[h]
	\centering
	\includegraphics[width=0.6\textwidth]{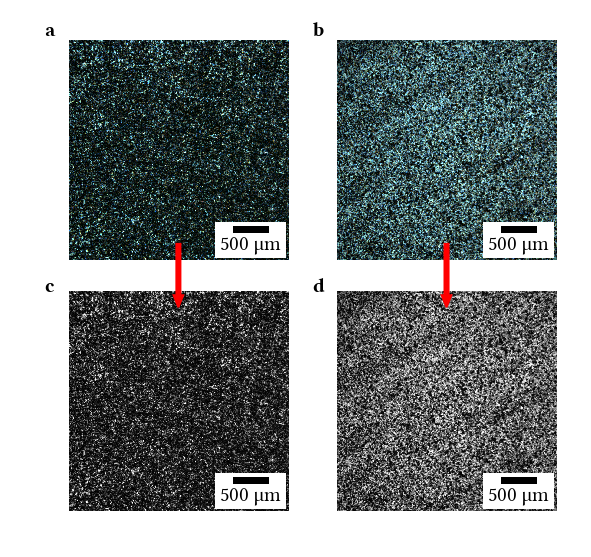}
	\caption{Images collected from cross-polarized microscopy where (a) and (b) show contrast-enhanced images of the film in parallel (a) and perpendicular (b) alignment orientations with respect to the incident polarization. The images are then converted to gray scale images (c) and (d) where an average pixel intensity is calculated and used to determine $I_{\mathrm{R}}$. 
	\label{fig:intensityRatio}}
\end{figure*}

The final characterization technique that we used was birefringence microscopy.
This method provides a very fast and simple characterization technique that can be used to accurately estimate \stwod .
The images acquired via this technique are shown in \autoref{fig:intensityRatio}.
These collected images (without contrast enhancement) are then converted to a gray scale image where pixel values range from 0 to 1.
Using these gray scale images, an average pixel intensity is then determined for the image from which a ratio, $I_{\mathrm{R}}$, can be calculated:
\begin{equation}
I_{\mathrm{R}} = 1 - \frac{I_{\mathrm{dark}}}{I_{\mathrm{bright}}},
\label{eq:intRatio}
\end{equation}
\noindent where $I_{\mathrm{bright}}$ corresponds to the average pixel intensity of the image collected from the HV case, and $I_{\mathrm{dark}}$ corresponds to the average pixel intensity of the image collected from a 45 degree rotation of the sample.

Using this intensity ratio, we are able to quickly make an estimation regarding the overall alignment of the film in both reflection and transmission.
The real strength of this method lies in the ability to do this without having to transfer the film to a substrate, i.e., while it is still on the membrane.
For the set of images shown in the figure, $I_{\mathrm{R}}$ was determined to be $0.71$, indicating that in this region of the film there is a relatively high degree of alignment.
For films that to not exhibit any alignment, $I_{\mathrm{R}}$ values are found to be well below $0.1$ pointing to the accuracy of this method. 

\newpage

\section{Flow rate control}
As shown in the main text, creating a constant flow rate increases SWCNT alignment using small diameter membranes.
Since different SWCNTs will pack differently into the film, and one of the main strengths of this method is that dispersions can be chirally separated prior to filtration, being able to control the flow rate without having to monitor drip rates allows for quick tuning of the flow rate to achieve high, global degrees of alignment.

To control the flow rate we first determine the resistance to filtration caused by the membrane, $R_\mathrm{m}$. To do this we start with Equation 1 in the main text:\cite{MulderMembrane1996}
\begin{equation}
\frac{1}{J}=\frac{\mu(R_\mathrm{m} + R_\mathrm{c})}{\Delta P}=\frac{\mu(R_\mathrm{m}+\alpha_{\mathrm{cake}} Cv)}{\Delta P},
 \label{eq:cakeFiltration}
\end{equation}
\noindent where $J$ is the permeation flux, $\mu$ is the viscosity of the permeate (SWCNT solution), $\alpha_{\mathrm{cake}}$ is specific cake resistance, $C$ is slurry concentration, $v$ is filtrate volume per unit area, and $\Delta P$ is the sum of applied pressure, $P_{\mathrm{applied}}$, and the head pressure, ${P_\mathrm{head}}$, which is equal to $\rho g h(t)$, where $\rho$, $g$, and $h(t)$ are the density, acceleration due to gravity, and the height of the meniscus, respectively.
The cake resistance, $R_\mathrm{c}$, is used as a substitution in this equation such that $R_\mathrm{c} = \alpha_{\mathrm{cake}} Cv$.

\begin{figure} [h]
	\centering
	\includegraphics[width=0.6\textwidth]{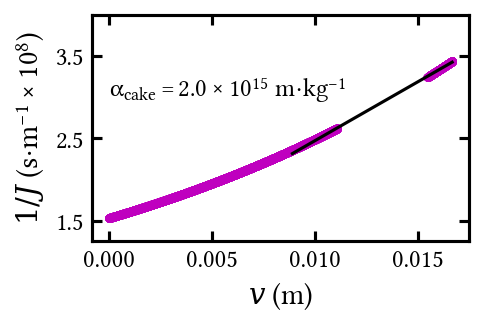}
	\caption{Determination of the specific cake resistance, $\alpha_{\mathrm{cake}}$, from the linear fit (black line) of the reciprocal of permeation flux, $J$, as a function of filtrate volume per unit area, $v$. This data was collected using a constant applied pressure and a SWCNT of 8~$\mu$g/mL and a DOC concentrations of 0.03~wt.\%.
\label{fig:cakeResistance}}
\end{figure}

To determine $R_\mathrm{m}$ (Figure 2b of the main text), we ran 4~mL of water through our filtration system at a constant pressure and room temperature.
For this condition, \autoref{eq:cakeFiltration} then becomes 
\begin{equation}
\frac{1}{J}=\frac{\mu R_m}{\Delta P},
\label{eq:rm}
\end{equation}
\noindent since no cake is being formed in this process $(R_{\rm c}= 0)$.
Data is collected by tracking the height of the meniscus, $h(t)$ as a function of time, which is used for determining the volume $\left[=h(t)\times A\right]$ and $P_\mathrm{head}$ $\left[\propto h(t)\right]$, where $A$ is the filter area (2.18~cm$^{2}$).

This process is repeated for a number of different $\Delta P$ values and then plotted, as shown in Figure 2b of the main text. From this data, we are able to extract the value of $R_\mathrm{m}$ for both the 80~nm and 200~nm pore size membranes. 

To determine the specific cake resistance, $\alpha_{\mathrm{cake}}$, a similar process is carried out. However, instead of using water, we use a SWCNT dispersion with a DOC concentration of 0.03~wt.\%, a SWCNT concentration of 8~$\mu$g/mL (estimated via optical density), and a constant applied pressure at room temperature. Due to the fact that we are at such low surfactant and SWCNT concentrations, we assume that the viscosity is very near that of water (8.9$\times 10^{-4}$~Pa$\cdot$s) at room temperature.

\begin{figure*} [h]
	\centering
	\includegraphics[width=0.97\textwidth]{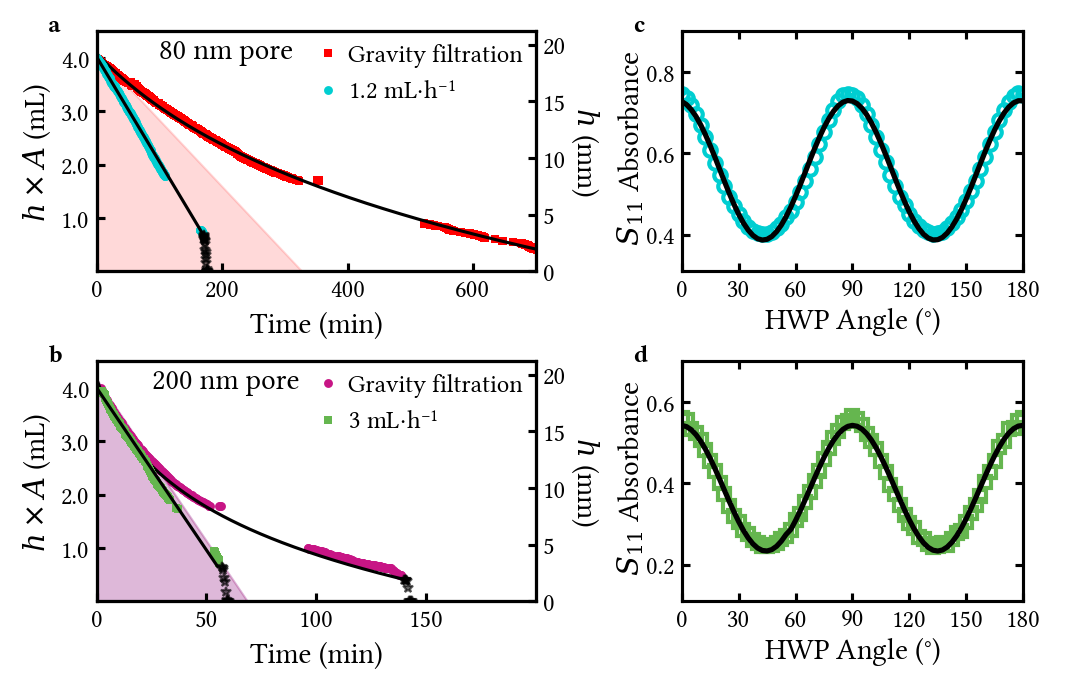}
	\caption{
	Data from Figure 2 of the main text for the (a) 80~nm and (b) 200~nm membrane pore size.
	The smaller pore size membrane has a higher $R_{\mathrm{m}}$, which allows us to explore a greater range of flow rates.
The shaded regions (pink and purple) again indicate flow rates that are accessible with the 80~nm pore size membrane and 200~nm, respectively. The breaks in the data in both (a) and (b) stem from the inability of the camera to track the meniscus edge at that location due to the lip of the filtration funnel. 
	(c) and (d) show polarized absorbance data of the \soneone\ peak from the controlled flow rate films made using the 80 nm and 200 nm filter membranes, respectively, created in (a) and (b).
\label{fig:flowrate}}
\end{figure*}

In this case, the collected data is plotted as the reciprocal of the permeation flux as a function of permeate volume per unit area, as shown in \autoref{fig:cakeResistance}. From \autoref{eq:cakeFiltration}, the slope of the best fit line is directly proportional to $\alpha_{\mathrm{cake}}$.
The fit is taken in the most linear region, near the end of the data set, since in the beginning of the filtration process there is little to no cake formed on the membrane.
Therefore, the cake filtration regime does not occur until nearly the end of the filtration process\cite{MulderMembrane1996}.

To the best of our knowledge, this value for $\alpha_{\mathrm{cake}}$ is the first report of specific cake resistance for SWCNTs.
Since no $\alpha_{\mathrm{cake}}$ values for SWCNTs are available, we use specific cake resistances from multi-wall carbon nanotubes (MWCNTs) as a means of comparison.
Interestingly, $\alpha_{\mathrm{cake}}^{\mathrm{SWCNT}}$ is an order of magnitude larger than $\alpha_{\mathrm{cake}}^{\mathrm{MWCNT}}$\cite{ZhangPorMat2014}, a difference we attribute to the high density packing of SWCNTs compared to MWCNTs afforded by the smaller SWCNT diameters.
The increased packing results in a low cake porosity\cite{MulderMembrane1996} and thus a higher $\alpha_{\mathrm{cake}}$ value.
Although this work is still ongoing, we have already observed that there is a significant change in $\alpha_{\mathrm{cake}}$ between rate zonal- and semiconductor-sorted SWCNTs, which we attribute as before to differences in average diameter.

Given the values found for $R_{\mathrm{m}}$ and $\alpha_{\mathrm{cake}}$, a desired $v$ as a function of time is then fed into the model where the necessary $P_{\mathrm{applied}}$ to achieve a constant $J$ is calculated. Experimental results from this method can be seen in Figures~\ref{fig:flowrate}a and b, where the model produces the desired flow rates of 1.2~mL$\cdot$h$^{-1}$ and 3~mL$\cdot$h$^{-1}$, respectively. 

\newpage

\section{Transfer of SWCNT films to a substrate}

\begin{figure*} [h]
	\centering
	\includegraphics[width=0.85\textwidth]{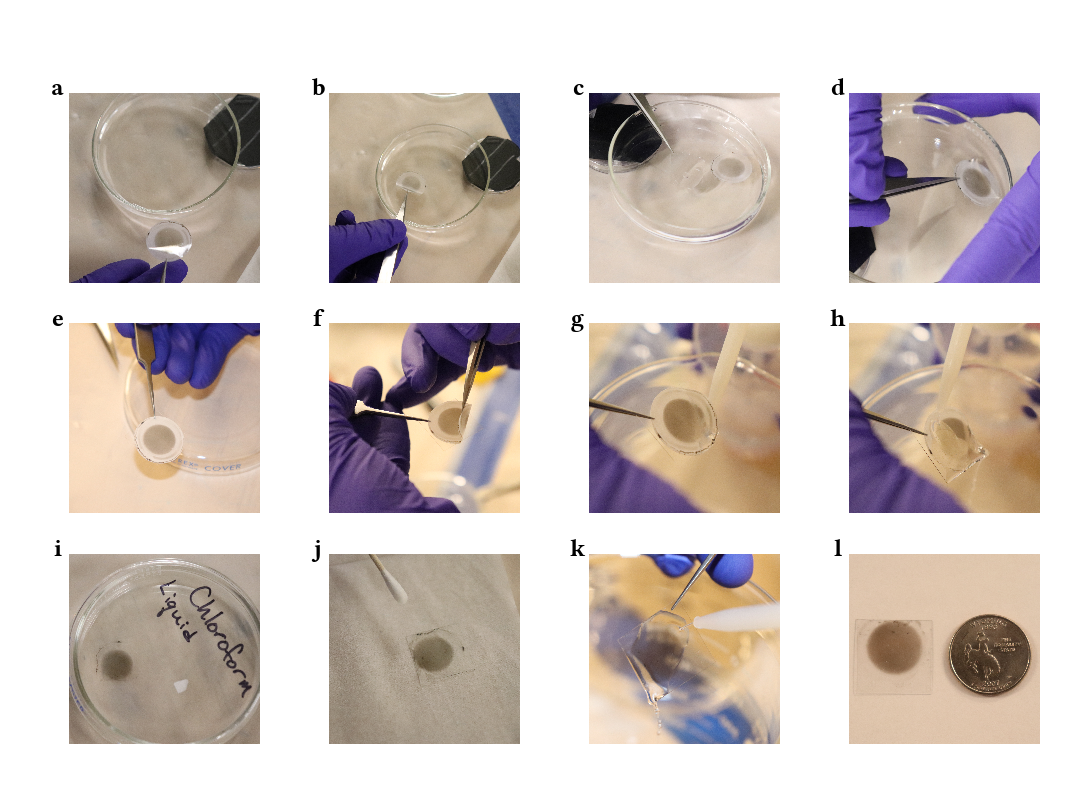}
	\caption{
	Photos describing the transfer process for SWCNT films. A film and membrane (film side up) (a) is placed film side down (b) on the surface of an \ce{H2O} bath. 
	A substrate is then placed on the bottom of the bath (c) and the film is floated onto the substrate surface.  Next, the film-substrate ensemble is removed from the bath (d).
	The film is then allowed to completely dry in air after which the film and membrane are adhered to the surface of the substrate (e).
	Next, a single corner is lifted from the substrate surface (f), leaving the majority of the film still adhered to the surface.
	Chloroform is then sprayed via a wash bottle under the film (g) permanently adhering the film to the substrate surface.
	The membrane is rinsed with chloroform (h) until the majority of the membrane has dissolved.
	The film and substrate are then placed in a liquid chloroform bath (i) at room temperature and left for $\approx$10~minutes.
	A DOC-wetted cotton swab is used to remove excess SWCNT from the outer edges of the substrate (j).
	A final chloroform rinse is used to remove any residual membrane (k), and the film is finally rinsed in acetone followed by nanopure \ce{H2O}.
	The transferred film is shown in (l). 
\label{fig:xferProcess}}
\end{figure*}

In order to make transmission optical measurements, we needed to transfer the film from the nanoporous membrane to an optically-transparent substrate.
Once the film has been dried with the accelerated flow rate, the film is allowed to finish drying in air until the residual liquid has evaporated, a process that typically takes a few hours (see \autoref{fig:xferProcess}a).
The film is then placed film side down on the surface of water in a Petri dish and allowed to float (\autoref{fig:xferProcess}b).
Depending on the measurement, a glass cover slip is used as the substrate and submerged in the water and placed under the film on the surface (\autoref{fig:xferProcess}c).
Prior to submersion, the cover slip is first rinsed in ethanol and then in filtered water.
The film is floated onto the surface of the substrate (\autoref{fig:xferProcess}d) and removed.
Excess water is shaken off the cover slip, and the film and cover slip are allowed to dry completely in air until the membrane has adhered itself to the glass surface.
This step can take anywhere from two to eight hours; in general, it is left overnight to dry.
A photo of the dry film can be seen in \autoref{fig:xferProcess}e.

A corner of the membrane is then slightly lifted, enough to get a wash bottle tip under the membrane.
A wash bottle tip with chloroform is then placed underneath the lifted corner and chloroform is lightly sprayed under the film (\autoref{fig:xferProcess}g).
This step permanently adheres the film to the glass surface.
Chloroform is then sprayed over the top of the membrane until the majority of the membrane has dissolved (\autoref{fig:xferProcess}h).
The film and substrate are then submerged in a liquid chloroform bath and soaked for $\approx$10~minutes and removed (\autoref{fig:xferProcess}i).
The sample is removed, and a cotton swab wetted with 4~wt.\% DOC is used to remove the excess SWCNT from around the film.
A final chloroform spray (\autoref{fig:xferProcess}k) is used to remove the remaining traces of the membrane.
The film is lightly rinsed with acetone to check for residual membrane and then finally rinsed with water and blow dried with dry air.
A transferred film created with this process is shown in \autoref{fig:xferProcess}l.
We note that sometimes residue (clumps) remain on the back of the substrate, which can be removed with a DOC-wetted cotton swab.

We find that using this reproducible method for transferring SWCNT films allows for the film to remain intact.
From other methods that have been attempted, it is very difficult to keep the film entirely adhered to the substrate in a reproducible manner.
While a risk still remains that the film detaches in places causing rips or holes in the films, we find that partial detachment is greatly reduced with our transfer method.

Other film transfer methods were also tested before we settled on the aforementioned procedure. For instance, when we submerged the films in either n-methylpyrrolidone (NMP) or chloroform, especially using glass substrates, we found that pockets of gas formed at the interface of the film and substrate.
After the membrane had been significantly removed, these pockets rupture leaving large gaps in the films, making a majority of the area unusable for further experiments.
Another group has reported using a combination of NMP vapor and heated NMP which may produce an equally robust transfer method~\cite{ChenACS2019}.
However, when we tried a process involving a similar combination of chloroform vapor and room temperature chloroform, we found the process to be unreliable.

The main difference we found between the use of NMP and chloroform was the rate at which the membrane dissolves.
In chloroform, the reaction is very fast, while in NMP (at room temperature), it is much slower.
It was also noted that using non-preserved chloroform can be catastrophic to the success of film transfer.
The reaction seems slightly slowed by the ethanol preservative found in most commercially-available chloroform, and this slowed rate seems to be beneficial in preserving the film during the transfer process.

\newpage

\section{Silane treatment of glassware}

In the main text, we discussed the implications of meniscus combing on the surface of the SWCNT film, which was shown to prevent global SWCNT alignment. 
We found that this happened in a number of different ways.
Depending on the surfactant concentration, as well as the final filtration speed, concentric rings would form, seemingly following the meniscus.
As the meniscus contacts the SWCNT film, a circle forms on the surface of the film where the liquid in the meniscus has already been pulled through the membrane and film leaving a visible, nearly dry area when taking a top-down perspective.
As filtering continues, this dry circular area grows in a radial manner, pulling the top layer of the SWCNT film with it.
This pinning of the meniscus to the SWCNT film is what we believe leads to the 2D spherulite formation.

\begin{figure*} [h]
	\centering
	\includegraphics[width=0.8\textwidth]{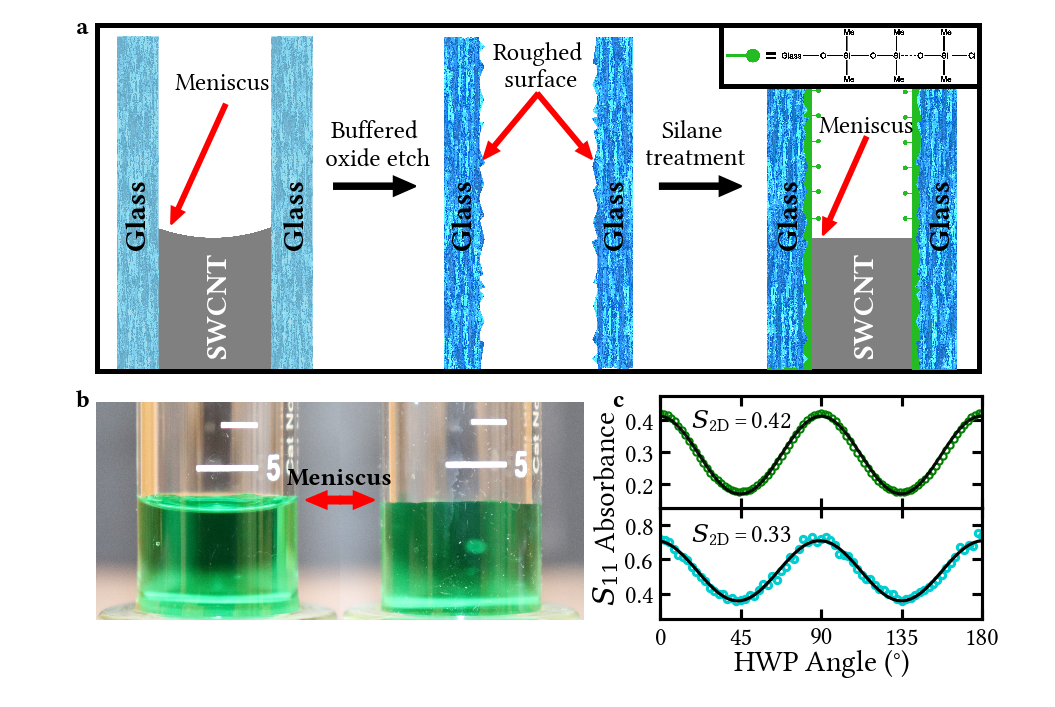}
	\caption{
	(a) From left to right, depiction of the silanation process of glassware. The filtration funnel prior to silane treatment with meniscus indicated. The untreated glassware is then etched using a buffered oxide etchant. Next, the glassware is reacted with silanization solution, which attaches silanol (inset) to the glass surface.
	A photo showing meniscus inside filtration funnel (b) before (left) and after (right) silane treatment where green food dye has been added to \ce{H2O} to act as a contrasting agent for the photos.
	(c) Polarized absorbance data taken at the $S_{11}$ peak of a film made with (top) and without (bottom) silane-treated glassware.
\label{fig:silaneTreatment}}
\end{figure*}

Many attempts at preventing meniscus combing were made including: increasing and decreasing the accelerated final flow rate, adjusting the initial temperature of the SWCNT dispersion (to either increase or decrease the viscosity of the dispersion), and changing the surfactant concentration (to again increase or decrease the viscosity of the dispersion).
We found that none of these methods produced adequate results and that either concentric rings formed in the film or a radial polarization of SWCNTs emerged in the upper film layer.

To eliminate this problem directly, we removed the concave meniscus all together from the glassware funnel.
Meniscus flattening was done via the use of a silane treatment of the funnel.
This methodology effectively coated the surface of the funnel with a short polymer of dimethylsiloxane and produced a super-hydrophobic surface inside the assembly funnel.
In \autoref{fig:silaneTreatment}a, we show a step-by-step schematic of the process, followed by pictures of the meniscus before and after the silane treatment (\autoref{fig:silaneTreatment}b).

The step-by-step procedure for silane treatment is as follows:\cite{SeedCellBio2000}

\textbf{Step I:} Clean the glass surface. We used an organic solvent (acetone) and \ce{H2O} to wash the glass surface.
After rinsing, dry air was used to mitigate any evaporative residue caused by the acetone.
The funnel was then rinsed thoroughly with nano-pure \ce{H2O} and again blow dried with dry air.

\textbf{Step II:} Chemically etch the glass surface. For etching, a buffered oxide etchant was used (10:1 ammonium diflouride) at room temperature.
A plastic beaker was placed on a stir plate with magnetic stir bar and the funnel was submerged for three minutes and removed, followed by an extensive water rinse.
This step roughens the glass surface at the micron level increasing the total surface area for a more effective silane treatment. 

\textbf{Step III:} Glass steam treatment.
A steam treatment was then used to incorporate more \ce{H2O} into the glass.
This process was performed using a hot plate, bell jar, and beaker of \ce{H2O}.
The beaker, along with the funnel, are placed on the hot plate and covered with the bell jar, while the beaker is in direct contact with the hot plate and the funnel is placed on an insulating block.
Heat is applied to boil the water and create a steam bath inside the jar.
500~mL of \ce{H2O} was used and heated until all the water had evaporated.
Since the \ce{H2O} molecules in the glass are one of the primary reactants in the process, we found this step to be helpful in obtaining a long-lasting silane treatment.
Once the water has completely evaporated, we allowed the funnel to completely cool back to room temperature before beginning the next step.

\textbf{Step IV:} Silane reaction.
In a fume hood, the glass funnel is placed inside a vacuum desiccator along with a cleaned petri dish with 3~mL of $\approx$5~wt./v\% dimethyldichlorosilane in heptane (\ce{C2H6Cl2Si}) solution.
The desiccator is then connected to a vacuum pump, and a vacuum is applied until the silane solution boils.
The chamber is sealed, the vacuum pump is removed, and the reaction is allowed to take place overnight.

\textbf{Step V:} Glassware final cleaning. After the reaction has had an ample amount of time to take place, the desiccator seal is broken and left open for a few minutes while remaining silane vapors evacuate the chamber.
Upon removing the funnel, it is rinsed in \ce{H2O} to remove any reactive chlorosilane ends of the polymer~\cite{SeedCellBio2000}.
The funnel is then cleaned with methanol and the process is complete. 

The difference in the meniscus is quite striking. As seen in \autoref{fig:silaneTreatment}b, the image of the meniscus in the funnel prior to silanation, and the meniscus in the same funnel taken after the silane treatment has been completed.
The volume in both images is equivalent (4~mL).
Food coloring was added to H$_2$O to provide contrast in the images. 

\newpage

\section{Spatially-resolved Raman mapping of the SWCNT films}

\begin{figure*} [h]
	\centering
	\includegraphics[width=0.8\textwidth]{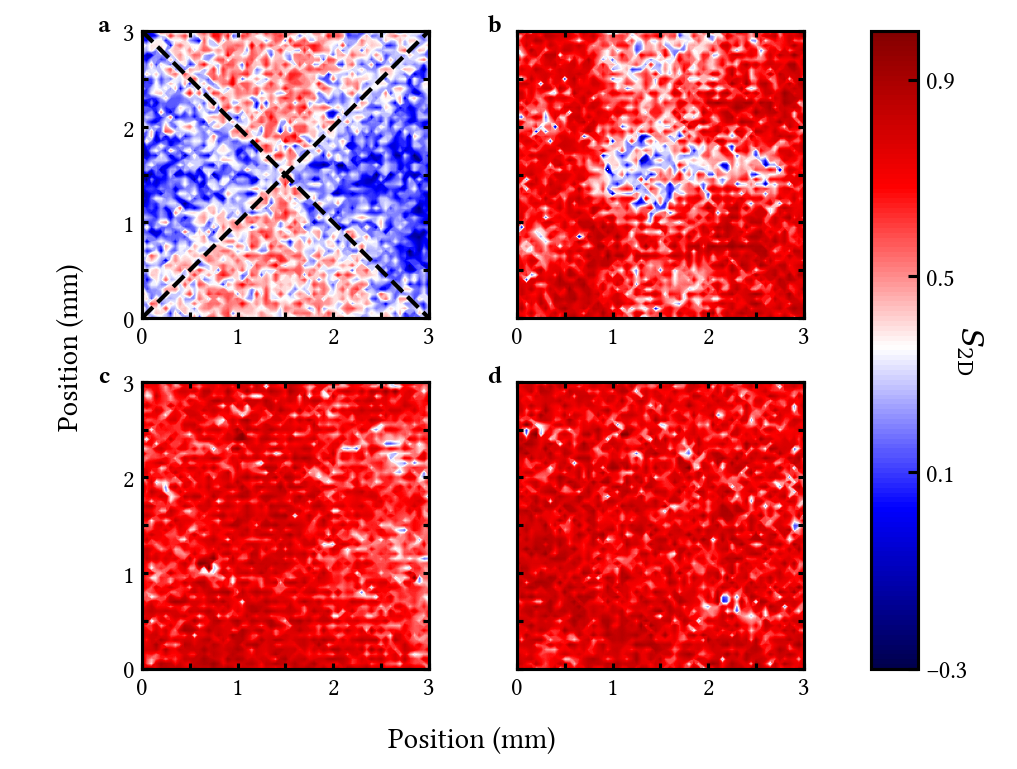}
	\caption{
	Spatially-resolved measurement of \stwod\ taken using polarized Raman spectroscopy. (a) and (b) are the back and front sides, respectively, of the SWCNT film made with non-silanated glassware. The radial polarization of the SWCNTs is not observed when measuring the front surface. (c) and (d) are the front and the back sides, respectively, of the SWCNT film made with silanated glassware. Global alignment is maintained on both film surfaces. (a) and (c) are shown in the main text.
These maps provide clear evidence that the 2D spherulite is removed through the silanation process, which leads to a high degree of global alignment. 
\label{fig:frontRamanMaps}}
\end{figure*}

As discussed in the main text, meniscus combing in non-silanated glassware affects the top side of the aligned SWCNT film.  Because of the film transfer process, this radial SWCNT alignment ends up on the back side of the SWCNT film when placed on a spectroscopically-appropriate substrate.  Here, we show that the radial alignment of the SWCNTs created by meniscus pinning does \textit{not} extend to the other side of the film.
In fact, as shown in \autoref{fig:frontRamanMaps}, 2D spherulite formation is \textit{only} observed on the side of the film that was in contact with the meniscus.
Critically, in the silanated case, the SWCNT alignment extends to \textit{both} sides of the film, which supports our claim of global alignment.

\begin{figure*} [h]
	\centering
	\includegraphics[width=0.8\textwidth]{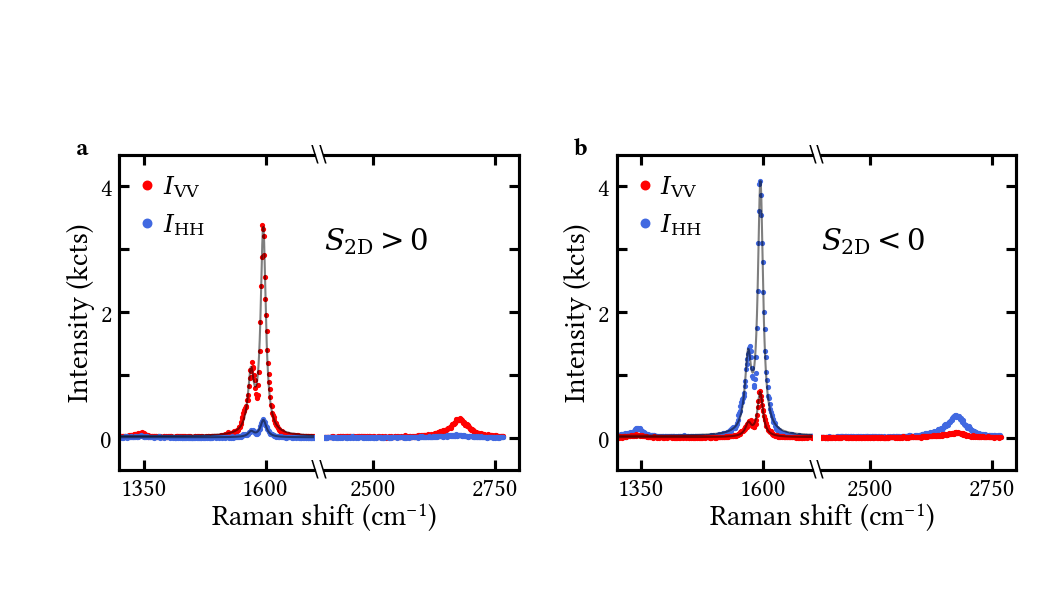}
	\caption{
	Individual Raman spectra extracted from Figure 3d of the main text showing the differences in regions where $S_{\rm 2D} > 0$ (a) and  $S_{\rm 2D} < 0$ (b). Black lines are fits to the G$^+$ and G$^-$ features.
\label{fig:RamanSpectra}}
\end{figure*}

In \autoref{fig:RamanSpectra}, individual spectra are extracted from the 2D mapping data shown in Figure~\ref{fig:frontRamanMaps}a from both the red and blue regions.
These spectra clearly demonstrate that the overall nematicity is not constant.
In fact, in the blue regions, the nematicity is nearly perpendicular to that of the nematicity of the red sections, further showing the importance of using large-area optical techniques for determining global alignment.  

\newpage

\section{Membrane pretreatment}

\begin{figure*} [h]
	\centering
	\includegraphics[width=0.8\textwidth]{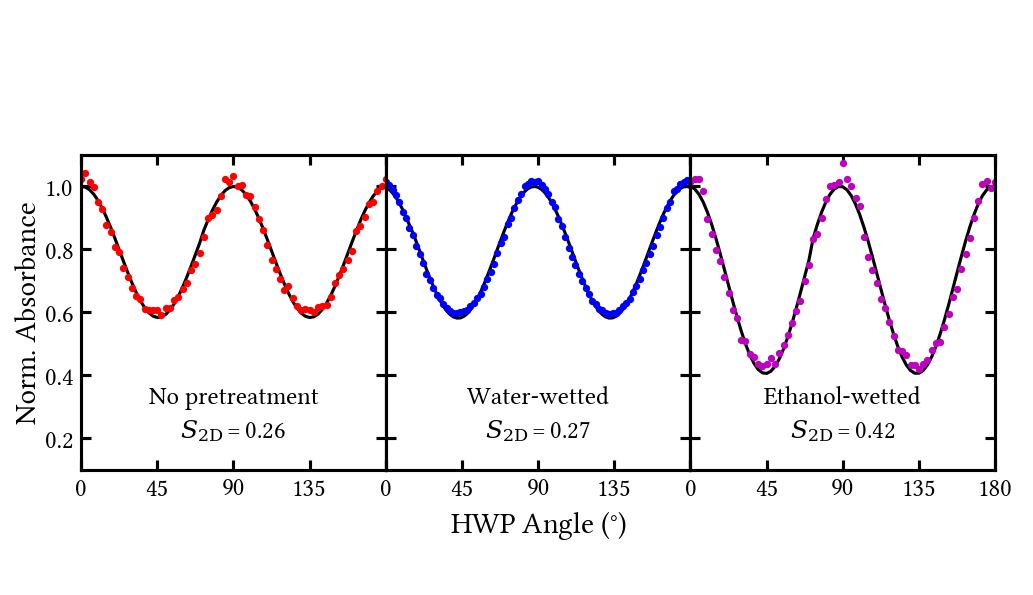}
	\caption{
Effect of dragging wipes across membrane surface prior to filtration. With no membrane pretreatment, we observe a low degree of alignment (left). When then using water-wetted wipe membrane treatment, we see a very similar degree of alignment (middle), suggesting that the effect of creating an added groove density (larger mechanical impact) to the membrane does not contribute to SWCNT alignment. However, when an ethanol-wetted wipe is dragged across the surface (right), there is a major uptick in the overall alignment of the SWCNT film, suggesting that directional tribocharging of the membrane drives the SWCNT alignment. Black lines are fits to the data.
\label{fig:membraneWiping}}
\end{figure*}

As can be seen in Figures 4a and b, the PVP-coated nanoporous membranes used for filtration possess a naturally-occurring grooved pattern.
It was first thought that these grooves may be contributing to the alignment, providing mechanical trenches in which the SWCNTs could fall into and provide the initial aligned layer.
However, these trenches are significantly larger than the SWCNTs, suggesting the mechanical effect of these grooves is much smaller than electrostatic forces.
To determine if this grooving had an effect on SWNCT alignment, we swept both a water- and ethanol-wetted wipe across the membrane surface along the direction parallel to the production-made grooves shown in Figure 4a.
As seen in \autoref{fig:membraneWiping}, despite the same mechanical action, it is only when the ethanol-wetted wipe is used that any increase in $S_{\rm 2D}$ is achieved.  
This observation strongly suggests that preferential electrostatic charging, instead of mechanical grooving, is responsible for SWCNT alignment using this filtration method.  

\newpage

\begin{mcitethebibliography}{61}
\bibitem[NIS()]{NISTDisclaimer}
Certain equipment, instruments or materials are identified in this paper in
  order to adequately specify the experimental details. Such identification
  does not imply recommendation by the National Institute of Standards and
  Technology (NIST) nor does it imply the materials are necessarily the best
  available for the purpose. \relax
\mciteBstWouldAddEndPunctfalse
\mciteSetBstMidEndSepPunct{\mcitedefaultmidpunct}
{}{\mcitedefaultseppunct}\relax
\EndOfBibitem
\end{mcitethebibliography}
\providecommand{\latin}[1]{#1}
\providecommand*\mcitethebibliography{\thebibliography}
\csname @ifundefined\endcsname{endmcitethebibliography}
  {\let\endmcitethebibliography\endthebibliography}{}

\end{document}